\documentclass [aps,twocolumn,superscriptaddress,altaffilletter,lengthcheck,tightenlines,showpacs,showkeys]{revtex4}
\usepackage[dvipdf]{epsfig}

\newcommand{\ben}{\begin{eqnarray}}
\newcommand{\een}{\end{eqnarray}}
\newcommand{\be}{\begin{equation}}
\newcommand{\ee}{\end{equation}}
\newcommand{\ba}{\begin{eqnarray}}
\newcommand{\ea}{\end{eqnarray}}

\begin{document}

%%%%%%%%%%%%%%%%%%%%%%%%%%%%%%%%%%%%%%%%%%%%%%%%%%%%%%%%%%%%%%%%%%%%%%%%%%%%%%%%%%%%%%%%%%%%%%%%%%%%%%%%%%%%%%%%%%%%%
\title{Cylindrical  wormholes in DGP gravity}

%%%%%%%%%%%%%%%%%%%%%%%%%%%%%%%%%%%%%%%%%%%%%%%%%%%%%%%%%%%%%%%%%%%%%%%%%%%%%%%%%%%%%%%%%%%%%%%%%%%%%%%%%%%%%%%%%%%%%

\author{Mart\'{\i}n G. Richarte}\email{martin@df.uba.ar}
\address{ Departamento de F\'{\i}sica, Facultad de Ciencias Exactas y
Naturales,  Universidad de Buenos Aires and IFIBA, CONICET, Ciudad
Universitaria, Pabell\'on I, 1428, Buenos Aires, Argentina}

\begin{abstract}
We construct traversable thin-shell wormholes in the Dvali-Gabadadze-Porrati   theory with cylindrical symmetry  applying the cut and paste procedure to a flat black string  solution of the five-dimensional vacuum Einstein field equations. In contrast to  general relativity case, where thin-shell wormholes violate both weak and null energy conditions, we  show that static  wormholes are supported by normal matter while  vacuum  wormholes do not exist.

%Besides, we show that the gravitational field associated to  these wormhole configurations is attractive for $m>0$.
\end{abstract}

\maketitle

%%%%%%%%%%%%%%%%%%%%%%%%%%%%%%%%%%%%%%%%%%%%%%%%%%%%%%%%%%%%%%%%%%%%%%%%%%%%%%%%%%%%%%%%%%%%%%%%%%%%%%%%%%%%
\section{Introduction}
%%%%%%%%%%%%%%%%%%%%%%%%%%%%%%%%%%%%%%%%%%%%%%%%%%%%%%%%%%%%%%%%%%%%%%%%%%%%%%%%%%%%%%%%%%%%%%%%%%%%%%%%%%%%
Alternative gravity theories such as one proposed by Dvali, Gabadadze, and Porrati (DGP) claims that  our observed four-dimensional universe lives in a larger five-dimensional space-time. In fact,  gravity is modified at large (rather than at short) distances through the slow evaporation of gravitational degrees of freedom of the brane universe \cite{dgp}.The transition between four and higher-dimensional gravitational potentials in the DGP model arises because of the presence of both the brane and bulk Hilbert-Einstein  terms in the action \cite{dgp}. Despite  the brane universe rendering the cosmic acceleration without need of any vacuum (dark energy) term \cite{ace1}, it suffers from some instabilities due to the existence of  ghost-like excitations \cite{dimitri}.

Any attempt to construct thin-shell wormholes requires the use of the cut and paste procedure \cite{visser1}, \cite{visser2} and work with the junction conditions associated to the gravity theory understudy \cite{daris}. Cylindrical thin shell wormholes within the context of general relativity (GR) were built, and found that, in most of the cases, the wormholes are supported by exotic matter, violating the energy conditions \cite{CWH}. Leaving aside the GR by taking into account  the Brans-Dicke gravity theory, it was shown that cylindrical thin-shells  are not necessarily are sourced by exotic matter and the energy conditions can be fulfilled  by  choosing  suitably the parameters of the model \cite{CWH2}.

In the case of  DGP-gravity theory, spherically symmetric thin-shell wormholes  have been constructed  by gluing two copies of the same five-dimensional vacuum solution (Schwarzschild black hole), where the throat of the wormhole is located at the joining surface \cite{dgpwh}. It turns out that the extra  Hilbert-Einstein term at the boundary has meaningful effects by generalizing the junction conditions \cite{jc2}; thereby it incorporates the Einstein tensor in the projected field equations on the joining surface, altering substantially the kind of matter which could support wormhole configurations. The contributions form the curvature tensor, theoretically, and seem to allow the existence of  wormholes supported by ordinary matter  as well as solitonic solutions, that is, vacuum shells  which are gravitationally self supported due to the nonlinear character of the junction conditions \cite{dgpwh}. %Furthermore,  the linear  stability  implemented for  
%traversable wormhole geometries supported by ordinary matter which satisfy the energy conditions, showed that there exist  stable wormholes with  squared speed sound  less then unity, indicating that the matter located at the throat  could be non-relativistic one \cite{dgpwh}. It is important to remark there are another kind of  alternative theories corresponding quadratic  corrections to the Einsten theory such as  the Einstein-Gauss-Bonnet model where it was found that spherical symmetric thin-shell wormholes could be supported by ordinary matter  fulfilling  the energy conditions \cite{mi1}. In the case of Brans-Dicke gravity, it was also claimed that, for certain negative values of the Brans-Dicke constant, the matter supporting the wormhole satisfies the energy conditions if the throat radius is suitably chosen \cite{mi2}. 

In the present work, we started by implementing the cut and paste procedure on the black string solution  in order to  construct static thin-shell wormholes with cylindrical symmetry, and then,  we  employed the generalized junction conditions within the DGP gravity to determine the energy density and pressure on the shell. We are going to address a key issue regarding the kind of matter which could support these wormholes configurations. We would like to know  whether  the DGP gravity provides a suitable framework to allow the existence of cylindrical wormholes which do not violate the energy conditions. In the next sections, without loss of generality, we will work with  units such that $8M^3_{5}=1$.

 %Later, we will investigate the  stability of these solution when the equations of state is selected.We will comment some interesting fact about the decoupling of the dynamics of the wormholes throat in that transverse section to the flat dimension.
%Finally, we will comment about the respulsive or attractive character of gravitational field for the cylindrical wormholes.

%%%%%%%%%%%%%%%%%%%%%%%%%%%%%%%%%%%%%%%%%%%%%%%%%%%%%%%%%%%%%%%%%%%%%%%%%%%%%%%%%%%%%%%%%%%%%%%%%%%%%%%%%%%%%%%%%%%%%%%%%%%%%%%%%%%%%%%%
\section{The DGP gravity }
%%%%%%%%%%%%%%%%%%%%%%%%%%%%%%%%%%%%%%%%%%%%%%%%%%%%%%%%%%%%%%%%%%%%%%%%%%%%%%%%%%%%%%%%%%%%%%%%%%%%%%%%%%%%%%%%%%%%%%%%%%%%%%%%%%%%%%%%%%
We begin by setting out the action corresponding to the DGP theory in a five-dimensional manifold ${\cal{M}}_5$  with a four-dimensional boundary $\Sigma$ (cf. \cite{jc2}),
\begin{eqnarray*}
\label{dgpa}
S&=&2M^{3}_{~5}\int_{{\cal M}_{5}} d^{5}{x}\sqrt{-g}R[g_{A\,B}] + 2M^{2}_{~4}\int_{\Sigma} d^{4}{x}\sqrt{-\gamma}{\cal R}[\gamma_{ab}] \\ 
&+&4M^{3}_{~5}\int_{\Sigma} d^{4}{x}\sqrt{-\gamma}\Big(-{\cal{K}}[\gamma_{ab}] + \frac{{\cal{L}}_{m}}{4M^{3}_{~5}}\Big),\\
\end{eqnarray*}
where $g_{A\,B}$ is the five-dimensional metric, $\gamma_{ab}$ is the four-dimensional induced metric on the boundary $\Sigma$, and $\cal{K}$ is the trace of extrinsic curvature. Here, the matter fields in ${\cal{L}}_{m}$ are confined to a 4-dimensional boundary. 
The extra term in the boundary introduces a mass scale $m_{c}=2M^{3}_{~5}/M^{2}_{~4}=r^{-1}_{c}$ which determines a scale that separates two different regimes of the theory. For distances much smaller than $m^{-1}_{c}$  one would expect the solutions to be well approximated by general relativity and the modifications to appear at larger distances. This is indeed the case for distributions of matter and radiation which are homogeneous and isotropic
at scales $\gtrsim r_{c}$. Typically, $m_{c}\simeq 10.42~ \mbox{GeV}$, so it sets the distance/time scale $r_{c}=m^{-1}_{c}$ at which the Newtonian potential significantly deviates from the conventional one. In the bulk the DGP  equations are the Einstein ones in vacuum : $G^{(5)}_{A\,B}=0$. For the coordinate  $X^{A}=(t,r,\theta, \phi,y)$ the bulk metric  corresponding to cylindrical black hole  vacuum solution takes the form

\begin{eqnarray}
\label{metric1} 
g_{A\,B}&=&\mbox{diag} \Big(-f(r), [f(r)]^{-1}, r^{2},r^{2} \mbox{sin}^{2}\theta,1\Big),
\\
\label{metric2} 
f(r)&=&\big( 1-\frac{r_{+}}{r} \big),
\end{eqnarray}
where the parameter $m$ is related to the  Arnowitt--Deser--Misner  mass. The above space-time has only one horizon placed at $r_{+}=2m$ with $m>0$. Besides, when $m<0$, the manifold only presents a naked singularity at the origin $r=0$ which can be easily verified through the squared Riemann tensor, given by ${R}_{A\,B\,C\,D}{R}^{A\,B\,C\,D}=10r^{2}_{~+}/r^{6}$ \cite{GLafla}. %It should be stressed that  $r$ corresponds to physical distance (radius) in three dimensions, so it does not contain a contribution from the flat fifth coordinate $y$.

%In order to get this  geometry Gregory and Laflamme extended four-dimensional Schwarzschild  black hole uniformly into the fifth dimension by adding of an extra flat coordinate therefore they found a cylindrical black hole which from now on we will call  black string solution \cite{GLafla}. Moreover they showed that the string vacuum solution, with the topology  $S_{\rm sch} \times R$, is clasically stable against linearized perturbation indicating that the topology of the horizon can be thought of as an extra balck hole hair\cite{GLafla}.

%In the next sections, without loss of generality, we will take $8M^3_{5}=1$ so the crossover scale reads as $r_{c}=4M^{2}_{~4}$. 

%%%%%%%%%%%%%%%%%%%%%%%%%%%%%%%%%%%%%%%%%%%%%%%%%%%%%%%%%%%%%%%%%%%%%%%%%%%%%%%%%%%%%%%%%%%%%%%%%%%%%%%%%%%%%%%%%%%%%%%%%%%%%%%%%%%%%%%%
\section{Thin-shell construction in DGP theory}
%%%%%%%%%%%%%%%%%%%%%%%%%%%%%%%%%%%%%%%%%%%%%%%%%%%%%%%%%%%%%%%%%%%%%%%%%%%%%%%%%%%%%%%%%%%%%%%%%%%%%%%%%%%%%%%%%%%%%%%%%%%%%%%%%%%%%%%%%%

Employing the metric Eqs.(\ref{metric1}-\ref{metric2}), we build a spherically symmetric thin-shell wormhole in DGP theory. We take two copies of the space-time and remove from each manifold the five-dimensional regions described by 
\begin{equation}
{\cal M}_{\pm}=\left\{x/r_{\pm}\leq a,a>r_{+}\right\}.
\end{equation} 
where $a$ is chosen to include possible singularities or horizon within the region ${\cal M}_{\pm}$. The resulting manifolds have boundaries given by the timelike hypersurfaces, 
\begin{equation}
\Sigma_{\pm}=\left\{x/r_{\pm} = a,a>r_{+}\right\}.
\end{equation}
Then we identify these two timelike hypersurfaces to obtain a geodesically complete new manifold ${\cal {M}}={\cal {M}}^{+}\cup {\cal {M}}^{-}$ with a matter shell at the surface $r=a$ where the throat of the wormhole is located. This manifold is constituted by two regions which are  asymptotically flat. To study this type of wormholes we apply the Darmois-Israel formalism generalized \cite{jc2} to the case of the  DGP theory.
We can introduce the coordinates $\xi^{a}=(\tau, \theta,\phi,y)$ in $\Sigma$, with  $\tau$ the proper time on the throat.  We will  focus in static configurations, then the  boundary hypersurface reads:
\begin{equation}
\Sigma: {\cal H}(r)=r-a=0.
\end{equation}
 
The  field equations projected on the shell $\Sigma$ are the generalized junction (or Darmois-Israel) conditions \cite{jc2}, 
\begin{equation}
\label{jc}
r_{c}[{\cal R}_{ab}-\frac{1}{2}\gamma_{ab}{\cal R}]-2\langle {\cal K}_{ab}-\gamma_{ab} {\cal K}\rangle={\cal S}_{ab},
\end{equation}
where the bracket $\left\langle .\right\rangle$ stands for  the jump of a given quantity across the  hypersurface $\Sigma$ and  $\gamma_{ab}$ is the induced metric on $\Sigma$. The extrinsic curvature ${\cal K}_{ab}$ is defined as follows:

\begin{equation}
{{\cal K}}^{\pm}_{ab}=-n^{\pm}_{A}\left(\frac{\partial^{2}X^{A}}{\partial\xi^{a}\partial\xi^{b}}+\Gamma^{A}_{B\,C}\frac{\partial X^{B}}{\partial\xi^{a}}\frac{\partial X^{C}}{\partial\xi^{b}}\right)_{r=a},
\end{equation} 
where $n^{\pm}_{A}$ are the unit normals  to the surface $\Sigma$.

Notice that the first term in (\ref{jc}) is not enclosed within the brackets because this contribution comes from the  four-dimensional Hilbert-Einstein term in the DGP action  which already lives in the boundary, so it does not need to be projected on $\Sigma$. By taking the limit  $r_{c}\rightarrow 0$,  we recover the standard Darmois--Israel junction condition found in \cite{daris}.  In order to proceed one can write the intrinsic metric to $\Sigma$ as 
\begin{equation} 
\label{Throat}
ds^{2}_{\Sigma}=-d\tau^{2} + a^{2}(d\theta^{2}+\mbox{sin}^{2}\theta d\phi^{2}) +dy^{2}.
\end{equation}
The position of the junction surface is given $X^{A}=(t(\tau), a, \theta, \phi, y)$ and the corresponding 4 velocity is $u^{A}=\big([f(a)]^{-1/2},0,0,0,0 \big)$, whereas the unit normal to shell may be determine by  the conditions $u^{A}n_{A}=0$ and $n^{B}n_{B}=1$. These requisites lead to the following expression, $n_{A}=\big(0,[f(a)]^{-1/2},0,0,0\big)$.
Noncompact wormhole  geometries such as the one described by  Eq. (\ref{Throat})  admit different variants of the definition of a throat \cite{Broni}. For instance,   the volume per unit length is given by  ${\cal V}(a)/\ell=4\pi a^{2}$ and  is an increasing function on both side of the throat.  Notice that  the throat has translational symmetry along the $y$direction, but when one   considers  $y$ fixed, the  global properties  are determined  by the behavior of topology of the  throat. So   our example corresponds to  a sphere with area  ${\cal A}(a)=4\pi a^{2}$, thus it has  a minimal ``area'' surface reaching  a minimum at the position of  the throat. Now, if we look at the topology of the throat for  $y$ and $\phi$ fixed,  we find that the circular radius function  ${\cal R}(a)=2\pi a$ defines its perimeter  and it can be considered as a less restrictive definition of the wormhole throat  \cite{Broni}.

Now, let us calculate some quantities that we will need later. The mixed components of the four-dimensional Einstein tensor are given by
\begin{eqnarray}
\label{ET} 
{\cal G}^{0}_{~0}&=&- \frac{1}{a^2}={\cal G}^{y}_{~y},
\\
\label{EE} 
{\cal G}^{\theta}_{~\theta}&=&={\cal G}^{\phi}_{~\phi}=0.
\end{eqnarray}
 The extrinsic curvature components read
\begin{eqnarray}
\label{KT} 
\left\langle {\cal K}^{0}_{~0} \right\rangle&=&\frac{ f'(a)}{\sqrt{f(a)}},\,\, \left\langle {\cal K}^{y}_{~y} \right\rangle=0.
\\
\label{KE} 
\left\langle {\cal K}^{\theta}_{~\theta}\right\rangle&=&\frac{2}{a}\sqrt{f(a)}=\left\langle {\cal K}^{\phi}_{~\phi}\right\rangle.
\end{eqnarray}
The component ${\cal K}^{y}_{~y}$ of the extrinsic curvature vanishes because the fifth dimension is flat, that is, this component involves the metric and its derivatives while ${\cal G}^{y}_{~y}$ is nonzero because  has a term proportional to the four-dimensional Ricci scalar $\cal{R}$. 

The  most general form of the stress energy tensor on shell compatible with the space-time symmetries is 
 \begin{equation}
 \label{tem}
 {\cal S}^{a}_{~b}=~\mbox{diag}~(-\sigma, P_{\theta}, P_{\theta}, P_{y})
\end{equation}
After some algebraic manipulation, we obtain that the energy density  and the tangential pressures  can be recast as
\begin{eqnarray}
\label{sigma} 
\sigma= \frac{r_{c}}{a^2}-\frac{8\sqrt{f(a)}}{a},
\\
\label{pe} 
P_{\theta}=
2\frac{f'(a)}{\sqrt{f(a)}}+\frac{4\sqrt{f(a)}}{a},
\\
\label{py} 
P_{y}=- \frac{r_{c}}{a^2}+
\frac{2f'(a)}{\sqrt{f(a)}}+\frac{8\sqrt{f(a)}}{a}.
\end{eqnarray}
The DGP contributions are encoded in the $r_{c}$ factor of the above equations. 

\section{Matter supporting the wormholes}
Classical solutions  within  the DGP model corresponding to thin-shell wormholes were found in \cite{dgpwh}, where the stability analysis indicated that these configurations  could be stable, moreover the matter supporting them can be interpreted as nonrelativistic in some cases due to their very small squared speed sound \cite{dgpwh}. All these elements seem to be  good reasons to consider a careful discussion about the nature of matter supporting wormholes  with cylindrical symmetry within the DGP model. Following the approach
presented  in \cite{dgpwh} where the four-dimensional Hilbert-Einstein generalizes the standard junction, adding  the Einstein tensor on the shell and due to its geometrical nature, the next approach will clearly be the most suitable framework to give a precise meaning to the characterization of matter supporting the wormhole with cylindrical symmetry.

The \emph{weak energy condition} (WEC) states that for any timelike vector $u^{A}$ it must be $T_{A\,B}u^{A}u^{B}\geq 0$;
the WEC also implies, by continuity, the \emph{null energy condition} (NEC), which means that for any null
vector $k^{A}$ it must be $T_{A\,B}k^{A}k^{B}\geq 0$ \cite{visser1}. In an orthonormal basis the WEC reads $\rho\geq 0$, $\rho + P_{l}\geq 0$ $\forall ~ l$  while the NEC takes the form $\rho + P_{l}\geq 0$ $\forall ~ l$.  %Besides, the \emph{strong energy condition} states that $\rho + P_{l}\geq 0$ $\forall ~ l$, and $\rho + 3P_{l}\geq 0$ $\forall ~ l$.

In the case of thin-shell wormholes the radial pressure $P_r$ is zero,  and within Einstein gravity, the surface energy density must fulfill $\sigma < 0$ so that both energy conditions will be violated. The sign of $\sigma+P_{t}$, where $P_t$ is the transverse pressure is not fixed, but it depends
on the values of the parameters of the system.

Now,  the sign of the surface energy density (\ref{sigma})  as well as the pressure  along axis $y$ (\ref{py}) is, in principle, not fixed. For $r_{c}\rightarrow 0$, we obtain the energy density and pressures for cylindrical wormhole  geometries  as if it was calculated  with the standard junction conditions.
Far away from the general relativity limit  we now find that there exist positive contributions to  $\sigma$. We stress
that this would not be possible if the standard Darmois-Israel formalism was applied; treating the
DGP contribution as an effective energy-momentum tensor,  thereupon we  inevitably would obtain that the energy density is  negative definite because the flare-out condition is fulfilled, so this leads to $\sigma=- 8[f(a)/a^{2}]^{1/2}<0$. %Although this identification is also possible; physically we would be  treating curvature objects  as an effective source for the junction condition. Moreover, based on effective energy-momentum tensor approach . 

Now, once the explicit form of the function $f(a)$ is introduced in Eq.($\ref{sigma}$), we focus on  the conditions  that lead to wormholes with $\sigma>0$. Then, it can be proven that wormholes with a non-negative surface density located at the shell are  allowable when the following inequalities are simultaneously satisfy: 

\begin{equation}
\label{snula2} 
\frac{r_{c}}{a^{2}}-\frac{8}{a} \Big(1-\frac{r_{+}}{a}\Big)^{\frac{1}{2}}>0\, \cap a-r_{+}>0,
\end{equation}
so it is always possible  to choose $a$ such that  the existence  of thin-shell wormholes is compatible with positive  surface energy density. More precisely. its radius  must belong to  the interval given below,
\begin{equation} 
 \label{int1}
 r_{+}<a\leq \frac{r_{+}}{2}+\frac{1}{2}\big(r^{2}_{+}+ \frac{r^{2}_{c}}{16}\big)^{\frac{1}{2}}
\end{equation}
Notice that the $r_{c}$-term is essential to have positive energy density; as one would expect, in the limit $r_{c} \longrightarrow 0$, this possibility completely vanishes. 
Besides, the sum of the pressure $P_{\theta}$ and energy density $\sigma$  takes the form
\begin{equation} 
 \label{int1}
 \sigma+ P_{\theta}=\left(\frac{r_{c}}{a^{2}}+\frac{2af'(a)-4f(a)}{a\sqrt{f(a)}}\right),
\end{equation}
because the first  term in (\ref{int1}) is positive the sign of $\sigma+ P_{\theta}$ depends on the second term, implying that the sum is positive  for $r_{+}<a\leq 3r_{+}/2$. Therefore, the remarkable result is that, now we have a region with $\sigma\geq0$ along with $\sigma+ P_{\theta}\geq0$. In addition, $\sigma+P_{y}=2 f'(a)/\sqrt{ f(a)}$ is always positive,  showing that wormholes with cylindrical symmetry within the DGP framework satisfy both  WEC and  NEC.

 We have obtained a nontrivial inequality (\ref{int1}) about the range that wormhole's radius must cover  in order to get  positive energy density. As is well known a physical  wormhole solution is  interestingly enough as long as the wormhole's radius it is not restricted to a small region in the parameter space defined by   $r_{c}$ and $r_{+}$. Therefore, it is important to check the nonexistence of such  fine-tuning between these two parameters, hence such analysis implies  to know  the order of magnitude of 
$r_{c}/r_{+}=4\pi r_{c} M^{2}_{~4}/m$, where the four-dimensional Newton constant is $G_{4}=1/8\pi M^{2}_{~4}$. The cosmological estimation of the crossover scale  using supernovae type Ia data leads to  $r_{c}=5 {\rm Gpc}$, whereas the parameter $m$ can be taken  as the black hole's mass. One could  infer the numerical value of $r_{c}/r_{+}=(1.33)\eta^{-1}\times 10^{24}$ by written the black hole mass in terms of the sun's mass provided $m=\eta M_{\odot}$, where $\eta$  encodes the size of black hole. 
For black holes  that are formed in the collapse of massive stars with stellar-mass ($\eta \simeq 10$) imply that the ratio $r_{c}/r_{+} \simeq {\cal O} (10^{23})$, whereas for supermassive black holes that reside  in galactic centers, this is tantamount to saying that  $10^{6} \leq \eta \leq 10^{10}$ \cite{PRL}; the ratio covers  interval $[10^{14}; 10^{18}]$. For instance, 
studying the dynamics of gas in the early-type galaxy NGC 4526, the statistical analysis leads to a central dark object of $4.5 \times 10^{8}M_{\odot}$ at $3\sigma$ confidence level \cite{Letter}. In this case, one obtains that $r_{c}/r_{+} \simeq 2.9 \times 10^{15}$.  Another appealing case seems to be  that of  black holes much less massive than a solar mass $M_{\odot}$ possibly formed as result of density perturbations in the early Universe. In the latter cases, the ratio is considerably amplified by many orders of magnitude provided that one always has $\eta \leq 10^{-1}$. In all the cases mentioned before, we  are able to ensure  that  $r_{c}/r_{+}$ is large enough to avoid any kinds of fine-tunings.

Finally, we would like to show that there are not vacuum wormholes with cylindrical symmetry self-gravitating due to the nonlinearity of  the junction conditions
within the framework of DGP gravity. In order to show that, we must impose that energy density and tangential pressures vanish at the same time. As it can be seen from Eq. (\ref{sigma}) and Eq.(\ref{py}), one can attain that $\sigma$ and $P_{y}$ vanish for a suitably radius, but Eq. (\ref{pe}) tells us that $P_{\theta}$ cannot become null; such a finding does not occur in the case of spherically symmetric wormholes \cite{dgpwh}.

%%%Agregar el onden de magintud como una discusion en las conlusiones y talvez estudiar la definicion de cosa areal y cosa ...

%%%%%%%%%%%%%%%%%%%%%%%%%%%%%%%%%%%%%%%%%%%%%%%%%%%%%%%%%%%%%%%%%%%%%%%%%%%%%%%%%%%%%%%%%%%%%%%%%%%%%%%%%%%%%%%%%%
\section{summary and discussion}
%%%%%%%%%%%%%%%%%%%%%%%%%%%%%%%%%%%%%%%%%%%%%%%%%%%%%%%%%%%%%%%%%%%%%%%%%%%%%%%%%%%%%%%%%%%%%%%%%%%%%%%%%%%%%%%%%%%%%%%%%%%%%%%%%%%%%%%%%%%%%%%%%%%%
In this work, we have taken a vacuum black string solution of the five-dimensional Einstein field equations and followed the cut and paste method for removing the singular part of this manifold in order to  construct a cylindrical wormhole. We have shown that wormholes with cylindrical symmetry within the framework of the DGP gravity theory do exist due to the nonlinear corrections that the boundary Hilbert-Einstein term adds to the generalized junction conditions.  

We have also proven that in a possible scenario, where the  DGP crossover scale $r_{c}$  is  considerably large enough in relation with the horizon radius $r_{+}$, corresponding to a situation far away from the general relativity limit, the energy density  located at the wormhole's throat can be positive; moreover, it turned out  that such gravitating configurations do fulfill the weak and null energy conditions. We have found that the  wormhole's radius is not restricted to a small region in the parameter space avoiding any kind of fine-tuning that could exist; thus, the DGP gravity theory  introduces a new
parameter, which allows for more freedom in the framework of determining the most viable wormhole configurations without to be threatened by the presence of exotic matter. For stellar black holes  with mass $m= 10M_{\odot}$, the ratio is $r_{c}/r_{+} \simeq {\cal O} (10^{23})$, whereas for supermassive black holes that reside  in galactic centers with  $10^{6} M_{\odot}\leq m \leq 10^{10} M_{\odot}$ \cite{PRL}, the ratio covers  interval $[10^{14}; 10^{18}]$. For instance, 
the statistical analysis performed to  the early-type galaxy NGC 4526 leads to a central dark object of $4.5 \times 10^{8}M_{\odot}$ at $3\sigma$ confidence level \cite{Letter}, implying  that $r_{c}/r_{+} \simeq 2.9 \times 10^{15}$.  Besides, one  difference in relation to the case with spherical symmetry is that now we cannot construct static vacuum wormholes; thus, we cannot have self-gravitating solutions  where both pressures and energy density 
vanish at the same time. Indeed,  such impossibility  arises  because  the anisotropic pressure along the axis where the black string is oriented in the original manifold does not vanish, however, we cannot discard that nonstatic vacuum wormholes do exist.

At this point,  one issue that remains open and  requires  further investigation is concerned with the dynamic of cylindrical wormholes in the DGP theory  (how these wormholes evolve when they are perturbated without altering their cylindrical symmetry).  In fact,  because  the black string background does not have the same horizon topology as the five-dimensional schwarzschild solution.  There is not a Birkhoff 's theorem that guarantees   no emission of gravitational waves (radiation), and hence the geometry outside the throat could not remain static \cite{GLafla}. This fact indicates that a more careful analysis must be carried out in treating cylindrical perturbations for wormhole geometries. 

%%%%%%%%%%%%%%%%%%%%%%%%%%%%%%%%%%%%%%%%%%%%%%%%%
\acknowledgments
%%%%%%%%%%%%%%%%%%%%%%%%%%%%%%%%%%%%%%%%%%%%%%%%%%%%%%%
We are grateful with the referee for his valuable comments that helped improve the article.

The author is partially supported by Postdoctoral
Fellowship Programme of  Consejo de Investigaciones Cient\'{\i}ficas y T\'ecnicas (CONICET).

\end{document}